\documentstyle[amssymb,prl,aps,twocolumn,epsfig]{revtex}

\def\R  {{\rm I\kern-.15em R}}   

\newcommand {\beq} {\begin{equation}}
\newcommand {\eeq} {\end{equation}}
\newcommand {\beqa}{\begin{eqnarray}}
\newcommand {\eeqa}{\end{eqnarray}}

\newcommand {\tr}{{\rm tr\,}}

\newcommand {\A}{ {\tilde A}}

\begin{document}
\draft

\title{
Dynamical generation of gauge groups\\
in the massive Yang-Mills-Chern-Simons matrix model
}

\author{Takehiro Azuma$^{1}$\cite{EmailTA},
        Subrata Bal$^{2}$\cite{EmailSB}
        and Jun Nishimura$^{1,3}$\cite{EmailJN}}
\address{$^{1}$
Theory Group, High Energy Accelerator Research Organization (KEK),\\
1-1 Oho, Tsukuba 305-0801, Japan \\
         $^{2}$
Theoretical Physics Laboratory, 
The Institute of Physical and Chemical Research (RIKEN),\\
2-1 Hirosawa, Wako, Saitama 351-0198, Japan \\
         $^{3}$Department of Particle and Nuclear Physics,
School of High Energy Accelerator Science,\\
Graduate University for Advanced Studies (SOKENDAI),
1-1 Oho, Tsukuba 305-0801, Japan}

\date{preprint  KEK-TH-1014, RIKEN-TH-41,
hep-th/0504217; 
April, 2005
     }
\twocolumn[\hsize\textwidth\columnwidth\hsize\csname@twocolumnfalse\endcsname

\maketitle
\begin{abstract}
\noindent
It has been known that the dynamics of
$k$ coincident D-branes in string theory
is described effectively by U($k$) Yang-Mills theory 
at low energy. While these configurations appear as classical solutions
in matrix models,
it was not clear whether it is possible to realize the $k\neq 1$ case
as the true vacuum.
The massive Yang-Mills-Chern-Simons matrix model 
has classical solutions corresponding to 
all the representations of the SU(2) algebra,
and provides an opportunity to address the above issue on a firm ground.
We investigate the phase structure of the model, and find in particular
that there exists a parameter region where 
O($N$) copies of the spin-1/2 representation appear as the true vacuum,
thus realizing a nontrivial gauge group dynamically.
Such configurations are analogous to the ones that are
interpreted in the BMN matrix model 
as coinciding transverse 5-branes in M-theory.
\end{abstract}
\pacs{PACS numbers 11.25.-w; 11.25.Sq}
]   

\paragraph*{Introduction.---}
The discovery of D-branes as classical solutions in string theory
suggested an interesting scenario that we are actually 
living on a kind of D-brane embedded
in a higher-dimensional space-time (``the brane-world scenario'').
The low energy effective theory of a D-brane is given
by a U(1) gauge theory, and in the case of $k$ coincident
D-branes, the gauge group enhances to U($k$).
Along this line, there are a lot of activities
in the search of a {\em perturbatively} stable 
brane configuration which realizes the Standard Model at low energy.

On the other hand, it is also possible
that our world 
is realized as the true {\em nonperturbative} 
vacuum of superstring/M theories.
Such an issue may be addressed by studying 
the matrix models \cite{9610043,9612115},
which are proposed as nonperturbative formulations of superstring/M
theories. 
In particular the IIB matrix model \cite{9612115}
can be obtained by dimensionally reducing
the 10-dimensional ${\cal N}=1$ super Yang-Mills theory
to a point. The space-time
should come out dynamically as the four extended directions
in the eigenvalue distribution of the 10 bosonic matrices \cite{Aoki:1998vn}.
The mechanism that realizes the brane world in this way
has been studied in Refs.\ \cite{NV,Anagnostopoulos:2001yb},
and by now there are certain evidences 
that indeed {\em four}-dimensional space-time is generated dynamically
\cite{0111102,0204240,0211272,0307007}.

The possibility of obtaining the U($k$) gauge group dynamically 
has been discussed in Ref.\ \cite{Iso:1999xs}.
There it was claimed that if the eigenvalues of the ten bosonic matrices
form clusters of size $k$,
the low energy
effective theory should have the U($k$) gauge symmetry. 
However, whether such clustering really occurs as a dynamical property of 
the IIB matrix model remains unclear.

In the series of papers \cite{0401038,0410263,0405277},
we discussed the same issue in matrix models
with a cubic term.
In these models the fuzzy spheres \cite{Madore} 
appear as classical solutions \cite{0101102} and play the role
of D-branes \cite{myers}. 
One of the advantages of 
studying fuzzy spheres in the present 
context is that they are solutions even for finite matrices,
and therefore one can study their dynamical properties by
well-defined perturbative calculations.
(See Refs.\ \cite{fuzzy} for other motivations for fuzzy spheres.)
The expansion around the $k$ coincident fuzzy spheres yields
a U$(k)$ gauge theory on a noncommutative geometry 
\cite{0101102}.
Therefore, by comparing the free energy for various fuzzy sphere
configurations, we may discuss the dynamical generation of gauge groups.
However, in the simplest Yang-Mills-Chern-Simons matrix
model \cite{0401038,0410263}, 
as well as in its higher-dimensional extensions \cite{0405277}, which
accommodate four-dimensional fuzzy manifolds,
the gauge group generated dynamically turned out to be U$(1)$.

In this letter we show that in fact one can obtain gauge groups of higher rank 
dynamically by adding a ``mass term'' \cite{0206075}
to the matrix models.
A similar model is known to appear in M-theory
on a plane wave background \cite{0202021}.
Although we consider that such a phenomenon occurs in more general models,
here we discuss the ``massive'' Yang-Mills-Chern-Simons matrix model 
for simplicity.

\paragraph*{The model and its classical solutions.---}
The model we consider in this letter is defined by the action
  \begin{eqnarray}
    S = N \, \alpha^4 
\tr \left\{ - \frac{1}{4} \, [A_{\mu} , A_{\nu}]^{2} 
  + \frac{2}{3} \,  i \,  \epsilon_{\mu \nu \lambda}
  A_{\mu} A_{\nu} A_{\lambda} \right. \nonumber \\
 \left. + \frac{1}{2} \, \rho^{2} \, (A_{\mu} )^{2}
  \right\} \ ,
 \label{action}
  \end{eqnarray}
where $A_{\mu}$ ($\mu=1,2,3$) are $N \times N$ Hermitian matrices.
(Rewriting the action in terms of 
${\cal A}_\mu = \alpha \, A_\mu$, one obtains
the model of Ref.\ \cite{0401038}
plus a mass term $\frac{1}{2} \, M^{2} \, ({\cal A}_{\mu} )^{2}$
with $M = \alpha \rho$.)
The classical equation of motion
 \begin{eqnarray}
   - [A_{\nu}, [A_{\mu}, A_{\nu}]] + i \,  \epsilon_{\mu \nu \lambda}
  \, [A_{\nu}, A_{\lambda}] + \rho^{2} A_{\mu} = 0
 \label{eom}
 \end{eqnarray}
has a class of perturbatively stable solutions given by
$ A_\mu = X_\mu \equiv \chi \, L_\mu $
for $0<\rho<\frac{1}{\sqrt{2}}$,
where $L_\mu$ ($\mu=1,2,3$) is an arbitrary $N$-dimensional
representation of the SU(2) algebra
and
$ \chi \equiv \frac{1}{2} \, \left( 1
  + \sqrt{1 - 2 \, \rho^2} \right) $.
Using the SU($N$) symmetry, the representation matrix $L_\mu$ 
can be brought into the block-diagonal form 
with $k_i$ blocks of the $n_i$-dimensional irreducible 
representation $L_\mu ^{(n_i)}$
satisfying $n_1 < n_2 < \cdots < n_s$ 
and $\sum_{i=1}^{s} n_i \, k_i = N$.
(The trivial solution $A_\mu = 0$ may be regarded as
a particular case in which one chooses $N$ copies of 
the representation $L_\mu^{(1)}=0$.)
Since the Casimir operator for each irreducible block
is given by
$ \Bigl( L_\mu^{(n)} \Bigr)^2  
= \frac{1}{4} \, (n^2 - 1)\, {\bf 1}_n $,
the configuration $X_\mu$
may be viewed as a collection of $k_i$ coincident fuzzy spheres
with the radius $\frac{1}{2}\, \chi \sqrt{(n_i)^2 -1}$,
and the classical action can be evaluated as
 \begin{eqnarray}
\label{cl-act}
 S_{\rm cl} &=& \frac{1}{4} \, N \, \alpha ^ 4 \, 
f(\chi) \, \sum_{i=1}^s k_i \, n_i \, 
\Bigl\{ (n_i)^2 - 1 \Bigr\} \ ,
 \end{eqnarray}
where $f(\chi) = \frac{1}{2} \, \chi^4  - \frac{2}{3}  \, \chi^3 
+ \frac{1}{2} \, \rho ^2 \, \chi^2 $.

Since $f(\chi)$ is positive (negative) for 
$\rho > \frac{2}{3}$ ($\rho < \frac{2}{3}$),
the minimum of $S_{\rm cl}$ is given by
the single fuzzy sphere $ A_\mu =  \chi \, L_\mu ^{(N)}$
for $\rho < \frac{2}{3}$, and
by the trivial solution $ A_\mu = 0$
for $\rho > \frac{2}{3}$.
These configurations describe the true vacuum in the large-$\alpha$
limit where quantum fluctuations are suppressed.
As we decrease $\alpha$, however, the one-loop effects become 
non-negligible, and it is possible that some other solution
describes the true vacuum.

\paragraph*{One-loop calculation.---} 

Let us evaluate the partition function
$Z = \int d A \, e^{-S}$ around a general solution
$A_\mu = X_\mu$ at the one-loop level.
The measure for the path integral is defined by
$dA =
 \prod_{\mu =1}^3 \prod_{a=1}^{N^2} \, dA_\mu^a $,
where $A_\mu = \sum_{a=1}^{N^2} A_\mu ^a \, t^a$
with $t^a$ being the generators of U($N$) normalized by
$\tr (t^a \, t^b)=\delta_{ab}$.

For solutions other than $A_\mu = 0$,
we need to fix the gauge since there are 
flat directions corresponding to the transformation
$A_\mu \mapsto A_\mu^{g} \equiv 
g \, A_\mu \,  g^\dag$, where $g$ is an element
of the coset space 
$H \equiv {\rm U}(N)/\prod_{i=1}^s [{\rm U}(k_i)]$.
We take the gauge fixing condition 
 \begin{eqnarray}
i \, [X_\mu , A _\mu] = C  \ ,
\label{gf-cond}
 \end{eqnarray}
where $C$ is a Hermitian matrix, and consider an identity
 \begin{eqnarray}
1 = \int_{H} 
dg \, \delta \, 
( \, i \, [X_\mu , A _\mu^{g}] - C \, ) \, \Delta(A_\mu) \ ,
\label{identity-rel}
 \end{eqnarray}
where the Faddeev-Popov determinant $\Delta(A_\mu)$ 
needs to be evaluated for $A_\mu = X_\mu$ at the
one-loop level.
Making an infinitesimal transformation $g=1+i \, h$ 
on $A_\mu = X_\mu$ in (\ref{gf-cond}), we obtain
$[ X_\mu , [X_\mu , h]] = C $.
 The linear transformation defined by the left hand side
has zero eigenvalues for $h$ belonging to the Lie algebra
of $\prod_{i=1}^{s} \mbox{U} (k_i)$.
We should therefore restrict the matrix $C$, as well as $h$,
to be within the tangent space of $H$. Since
$\Delta(X_\mu)$ is given by the product of the eigenvalues
of the above linear transformation in the restricted space,
we obtain
 \begin{eqnarray}
\Delta(X_\mu) &=& 
\prod _{i,j=1}^s 
\prod  _{l=|n_i-n_j|/2}^{(n_i+n_j)/2-1} \hspace{-6mm} {\bf '}
\hspace{6mm}
\Bigl[ \chi^2 \, l \, (l+1) \Bigr]^{k_i k_j (2l+1)}  \ ,
 \end{eqnarray}
where the symbol $\prod \, '$ implies that $l=0$ is excluded.
Inserting the unity (\ref{identity-rel}) and exploiting the
SU($N$) invariance, we rewrite the partition function as
 \begin{eqnarray}
Z =  \mbox{vol}(H) \!
\int \! dA \,  \delta
( i [X_\mu , A _\mu] - C ) \,
\Delta(A_\mu) \, e^{-S} \ ,
 \end{eqnarray}
where 
$\mbox{vol}(H)= \mbox{vol}({\rm U}(N))/ 
\prod_{i=1}^s \mbox{vol}({\rm U}(k_i))$
can be obtained by
$\mbox{vol}({\rm U}(n)) = 
\frac{(2 \pi)^{\frac{n(n+1)}{2}}}{(n-1)\, ! \cdots 0 \, !  }$ .
Since the result does not depend on $C$, we integrate over it
within the tangent space of $H$ with the Gaussian weight
${\cal N}
e^{-\frac{1}{2} N \alpha^4 \tr C^2 } $,
where 
${\cal N}
= \left( \frac{N\alpha^4}{2\pi} \right)
^{\frac{1}{2} \{ N^2 - \sum_{i=1}^s (k_i)^2 \} }$ .
This yields an extra term
$  S_{\rm g.f.} = - \frac{1}{2} \, N \, \alpha ^4 \,
  \tr [X_{\mu}, A_{\mu}]^{2} 
$
in the action, which lifts the flat directions as desired.

Now we are ready to perform the integration over $A_\mu$.
Decomposing the variables as $A_{\mu} = X_{\mu} + \A_{\mu}$,
we expand the total action $S_{\rm tot} = S + S_{\rm g.f.}$
with respect to the fluctuation $\A_{\mu}$ as
\begin{eqnarray}
 S_{\rm tot} = S_{\rm cl} +
\frac{1}{2} \, 
N \, \alpha^4  \, 
 \tr \Bigl(
\A_{\mu} {\cal Q}_{\mu \nu} \A_{\nu}
\Bigr) + \cdots  \ , \label{kinetic}
\end{eqnarray}
where the operator ${\cal Q}_{\mu \nu}$ is given by
\begin{eqnarray}
 {\cal Q}_{\mu \nu} = \Bigl\{ 
\chi^{2} \, ({\cal L}_{\lambda})^{2}+ \rho^{2} \Bigr\} 
 \,  \delta_{\mu \nu} - 
  i \, \rho^{2} \, \epsilon_{\mu \nu \lambda} \, {\cal L}_{\lambda} \ .
\end{eqnarray}
We have introduced an operator ${\cal L}_{\lambda}$ which
acts on a $N \times N$ matrix $M$ as
${\cal L}_{\lambda} \, M = [ L_{\lambda} , M] $.
The remaining task is to solve 
the eigenvalue problem 
${\cal Q}_{\mu \nu} \A_{\nu} = \lambda \, \A_{\mu}$.
Since $L_{\mu}$ has a block-diagonal form,
we can decompose the matrix $\A_{\mu}$ 
into blocks of size $n_i \times n_j$,
and it suffices to solve the eigenvalue problem 
within each block \cite{0412052}.
This can be done in a similar way as in the 
BMN matrix model \cite{0205185}.
As a complete basis for each block, we choose the eigenstates 
$| \, l,m  \rangle $ of $({\cal L}_{\lambda})^2$ and ${\cal L}_3$
with the eigenvalues $l \, (l+1)$ and $m$,
where $|m|\le l$ and $|n_i-n_j|/2 \le l \le {(n_i+n_j)/2-1}$.
Expanding each block of $\A_{\mu}$ as 
\begin{eqnarray}
(\A_{\mu})_{n_i \times n_j} 
= \sum_{l=|n_i-n_j|/2}^{(n_i+n_j)/2-1}
 \sum_{m=-l}^{l} \A_{\mu}^{(lm)} \, | \, l,m \rangle
\end{eqnarray}
and using the properties
\begin{eqnarray}
({\cal L}_1 \pm i {\cal L}_2) \, | \, l,m \rangle 
&=& b_{\pm} \, | \, l,m\pm 1 \rangle  \ , \\
b_{\pm} &=& \sqrt{ \, l \, (l+1) - m\, (m \pm 1)  }  \ ,
\end{eqnarray}
we obtain
the eigenvalue equation within each block as 
\begin{eqnarray}
\label{eig-eq1}
&~& \{  \Lambda - \rho^2  (1-m) \}  \A_{+}^{(l,m)} 
 =  \rho^2  b_{-}  \A_{3}^{(l,m-1)} \ , \\
&~& \{ \Lambda - \rho^2  (1+m) \}  \A_{-}^{(l,m)} 
 =  - \rho^2  b_{+}  \A_{3}^{(l,m+1)} \ , \label{eig-eq2} \\
&~&  \{ \Lambda - \rho^2 \}  \A_{3}^{(l,m)} 
  = \frac{1}{2}  \rho^{2}  ( b_{+}  \A_{+}^{(l,m+1)}
- b_{-}  \A_{-}^{(l,m-1)} ) \ , 
  \label{eig-eq3}
\end{eqnarray}
where we have introduced
$\A_{\pm}^{(l,m)} = \A_{1}^{(l,m)} \pm i \, \A_{2}^{(l,m)}$
and $\Lambda = \lambda - \chi^2  \, l\, (l+1)$.
For $l \ge 1/2$ we have eigenvalues
\begin{eqnarray}
\lambda_1 (l)&=& \chi^{2}\, l \,(l+1) \ , \label{sol1} \\
\lambda_2 (l)&=&  \chi^{2}\, l \, (l+1) - \rho^{2}\, l  \ , \label{sol2} \\
\lambda_3 (l)&=&  \chi^{2}\, l \,(l+1) + \rho^{2} \, (l+1) \ , \label{sol3}
\end{eqnarray}
whose degeneracy is $2 \, l+1$, $2 \, l-1$, $2 \, l+3$, respectively.
(For $l=1/2$ the 2nd eigenvalue does not appear.)
For $l=0$ we have $\lambda = \rho^2$ with 3-fold degeneracy.
Thus the integration over $\A_\mu$ yields
\begin{eqnarray}
\label{calZ}
{\cal Z} &=& 
\left( \frac{2\pi}{N\alpha^4} \right)
^{\frac{3}{2} \, N^2} \cdot
\prod_{i=1}^s
\rho ^{-3 (k_i)^2} \cdot 
\prod_{i,j=1}^s (q_{n_i n_j})^{k_i k_j}  \ , \\
q_{nm} &=&
\prod_{l=|n-m|/2}^{(n+m)/2-1}
\hspace{-6mm} {\bf '} \hspace{6mm}
\Bigl[ \lambda_1(l)^{2l+1}
\lambda_2(l)^{2l-1}
\lambda_3(l)^{2l+3} \Bigr]^{-\frac{1}{2}} \ .
\end{eqnarray}

Bringing all the factors together, we get
\begin{eqnarray}
Z =  \mbox{vol}(H) \,  {\cal N} \, 
\Delta(X_\mu) \,  {\cal Z} \,  e^{-S_{\rm cl}} \ .
\label{pfn-explicit}
\end{eqnarray}
In fact this result holds also for the case $X_\mu = 0$ formally.

If we take the large-$N$ limit of
the free energy $F = - \log Z$
with fixed $\alpha$ and $\rho$,
the one-loop terms give maximally
O($N^2 \log N$) contributions.
The O($N^2 \log N$) terms coming from 
$\mbox{vol}(H)$ and ${\cal N}$ cancel each other in general.
Apart from the universal term $\frac{3}{2}N^2 \log N$ coming
from the first factor in (\ref{calZ}),
we have contributions from $\Delta(X_\mu)$
and the third factor in (\ref{calZ}), which
induce extra O($N^2 \log N$) terms when $n_i$ becomes of O($N$).

\paragraph*{The $\rho < \frac{2}{3}$ regime.---}
Since the single fuzzy sphere 
has a {\em negative} classical action of O($\alpha^4 N^4$)
in this regime, no other solutions can be the true vacuum
unless $\alpha$ becomes as small as O($\frac{1}{\sqrt{N}}$).
The free energy for the single fuzzy sphere 
$A_\mu = \chi \, L_\mu^{(N)}$ is obtained as
\begin{eqnarray}
\frac{1}{N^2}  F_{\rm FS} = \frac{1}{4}\, N^2  \alpha ^4  f(\chi)
+ \frac{5}{2} \log N + 4 \, \log \alpha - \delta
\label{singleFS}
\end{eqnarray}
at large $N$, where $\delta$ is an O(1) constant which depends only
on $\rho$. (The one-loop calculation
is reliable for the single fuzzy sphere as far as
$\alpha\sqrt{N}$ is large \cite{0401038,0410263}.
This is not the case for all the solutions.)

On the other hand, at small $\alpha$
we find, after rescaling ${\cal A}_\mu = \alpha A_\mu$,
that the quartic term in the action (\ref{action})
becomes dominant, and obtain the so-called Yang-Mills phase \cite{0401038},
which is described
by the vacuum of the pure Yang-Mills model \cite{HNT}.
Using the result of 
the Gaussian expansion method \cite{Nishimura:2002va},
we obtain the free energy of this phase as
\begin{eqnarray}
\frac{1}{N^2}  F_{\rm YM} = \frac{3}{2} \,  \log N + 3 \, \log \alpha
+\gamma 
\label{f-YMphase}
\end{eqnarray}
at large $N$, where $\gamma  \sim -4.5$.

Comparing (\ref{singleFS}) and (\ref{f-YMphase}), we find that
$F_{\rm YM}$ becomes smaller than $F_{\rm FS}$ at 
\begin{eqnarray}
\alpha < \alpha _{\rm cr} \equiv
\frac{1}{\sqrt{N}} \left( 
\frac{2 \, \log N}{|f(\chi)|}
\right)^{1/4} \ .
\label{alpha-crit}
\end{eqnarray}
This is consistent with the Monte Carlo simulations at $\rho=0$
\cite{0401038},
where a first order phase transition has been observed
with the upper and lower critical points
$\alpha_{\rm cr}^{\rm (u)}\sim 0.66$ and
$\alpha_{\rm cr}^{\rm (l)} \sim \frac{2.1}{\sqrt{N}}$, 
respectively. Notice the inequality
$\alpha_{\rm cr}^{\rm (l)} < \alpha _{\rm cr}(\rho = 0)
< \alpha_{\rm cr}^{\rm (u)}$.

\paragraph*{The $\rho > \frac{2}{3}$ regime.---}
Since $f(\chi)>0$ in this regime, 
in order for some nontrivial solution
to have smaller free energy than the $A_\mu = 0$ solution,
it should have a classical action 
of O($N^2$) or of smaller order. Thus it turns out to be sufficient
to consider the case where $s$ and all the $n_i$
are of O(1), while $k_i$
are of O($N$). 
Such a configuration (except $A_\mu=0$)
is analogous to the ones that are interpreted 
\cite{Maldacena:2002rb} in the BMN model 
as a collection of coincident transverse 
5-branes in M-theory.
Let us denote $k_i = r_i \, N$, where $r_i$ are real parameters of O(1).
In what follows we consider the large-$N$ limit taken in this way.

Let us first consider the case where $\alpha$ is very large, 
so that we can neglect all the $\alpha$-independent terms 
in the free energy and obtain the relevant terms
\begin{eqnarray}
\frac{1}{N^2} F \sim   A \sum_{i=1}^s r_i \, n_i 
\Bigl\{ (n_i)^2 - 1 \Bigr\} 
+ B \sum_{i=1}^s (r_i)^2  \ ,
\end{eqnarray}
where
$A = \frac{1}{4} \, \alpha^4 \, f(\chi)$ and
$B = 2 \, \log \alpha $.
We determine the $r_i$ that minimize the free energy
within the constraint $\sum_{i=1}^s n_i r_i = 1$
using the Lagrangian multiplier $\lambda$ as
\begin{eqnarray}
r_i = \frac{1}{2B} \, 
\Bigl[ \lambda \, n_i - A \, n_i \, \Bigl\{ (n_i)^2 - 1 \Bigr\} \Bigr] 
\ ,
\label{proportion}
\end{eqnarray}
where $\lambda$ should be fixed by the constraint.
We may assume $n_p = p$, and requiring $r_i \ge 0 $, we obtain
$s(s^2-1)(4 s^2-1) < 60 B/A$,
where $s$ should be taken to be
the largest possible integer.

Thus
we obtain the $A_\mu = 0$ solution in the large-$\alpha$ limit
as expected, but as we go below the
critical $\alpha$ determined by $\frac{A}{B}=\frac{2}{3}$,
O($N$) numbers of $2\times 2$ blocks
start to appear.
If we decrease $\alpha$ further, 
larger and larger blocks
appear with the specific proportion (\ref{proportion}).
This result is valid when $\rho$ is very close to $\frac{2}{3}$
so that the coefficient $f(\chi)$ in $A$ is small
and the transitions take place at large $\alpha$.

\begin{figure}[htbp]
  \begin{center}
 \mbox{\epsfig{file=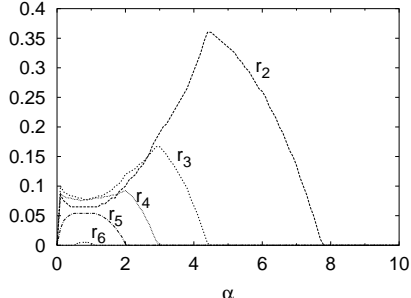,width=60mm}}
    \caption{The parameters $r_i$ ($2 \le i \le 6$)
that minimize the free energy are plotted against $\alpha$ 
for $\rho =0.7$. 
}
    \label{fig:r-rho0-700}
  \end{center}
\end{figure}

\begin{figure}[htbp]
  \begin{center}
 \mbox{\epsfig{file=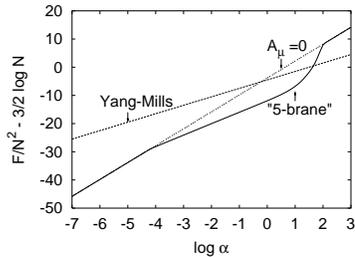,width=50mm}}
    \caption{The ``free energy density'' 
$\lim_{N\rightarrow \infty} (\frac{1}{N^2}F - \frac{3}{2}\log N)$
is plotted against $\log \alpha$ 
for $\rho =0.7$. The solid line represents the result obtained
by the one-loop calculation, which is valid at large $\alpha$.
The dip corresponds to some ``5-brane'' solution, 
and the straight part corresponds to the $A_\mu=0$ solution.
The dashed line represents
the result (\ref{f-YMphase}) obtained for the Yang-Mills phase,
which is valid at small $\alpha$.
}
    \label{fig:free-r-0-700}
  \end{center}
\end{figure}

At moderate $\alpha$, the one-loop terms
which are independent of $\alpha$ can no longer be neglected.
We search for the minimum of the free energy numerically
restricting ourselves to the case 
$n_p = p$ ($p = 1, \cdots , s $) with $s=7$.
In Fig.\ \ref{fig:r-rho0-700} we plot the result for 
$\rho = 0.7$. 
(The situation for other values of $\rho \, (>\frac{2}{3})$ 
is qualitatively the same.)
Since we obtain $r_7=0$ throughout the whole region of $\alpha$,
the above restriction is considered to be harmless.
The minimum free energy obtained in this way is plotted
in Fig.\ \ref{fig:free-r-0-700}, where we also plot
the result (\ref{f-YMphase}) for the Yang-Mills phase, 
which should be valid for small $\alpha$, and expected to be 
smoothly connected to the $A_\mu = 0$ solution at $\alpha \sim \mbox{O}(1)$.
In the $\rho > \frac{2}{3}$ regime, since
the action (\ref{action}) is positive
definite, we can easily prove that
the free energy should be a monotonously increasing (continuous)
function of $\alpha$.
We therefore consider that higher loop corrections increase
the free energy for the ``5-brane'' at $\alpha \sim \mbox{O}(1)$, 
where the Yang-Mills phase should take over.

\begin{figure}[htbp]
  \begin{center}
 \mbox{\epsfig{file=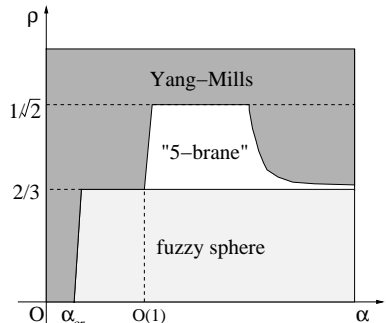,width=50mm}}
    \caption{A schematic view of the phase diagram of the massive
Yang-Mills-Chern-Simons model. The critical point 
$\alpha_{\rm cr}$ between two phases below $\rho = \frac{2}{3}$
is given by eq.\ (\ref{alpha-crit}). 
}
    \label{fig:phase-diagram}
  \end{center}
\end{figure}

Combining all the results obtained above, we arrive at 
the phase diagram depicted in Fig.\ \ref{fig:phase-diagram}.
\vspace*{-1cm}
\vspace*{-1cm}
%
%
%

\paragraph*{Acknowledgments.---} 
We would like to thank
H.\ Aoki, Y.\ Kitazawa, T.\ Kuroki, 
S.\ Mizoguchi, A.\ Mukherjee, K.\ Nagao, 
G.W.\ Semenoff, B.\ Ydri and K.\ Yoshida for valuable discussions.
The work of T.A.\ and J.N.\ is supported in part by Grant-in-Aid for 
Scientific Research (Nos.\ 03740 and 14740163)
from the Ministry of Education, Culture, Sports, Science and Technology.

\end{document}